\documentclass[useAMS,usenatbib]{mn2e}
\input epsf.sty

\title[Evolution of a neutrino-cooled disc in Gamma-Ray Bursts]{Evolution of a neutrino-cooled disc in Gamma-Ray Bursts}
\author[A. Janiuk, R. Perna, T. Di Matteo,  B. Czerny]{A. Janiuk$^{1, 
3}$\thanks{E-mail:
agnes@camk.edu.pl}, R. Perna$^{2}$, T. Di Matteo$^{3}$, B. Czerny$^{1}$ 
\\
$^{1}$Nicolaus Copernicus Astronomical Center, Bartycka 18,
            00-716 Warsaw, Poland\\
$^{2}$Princeton University, 4 Ivy Lane, Princeton, NJ 08542, USA\\
$^{3}$Max-Planck Institute f\"ur Astrophysik,
            Karl-Schwarzschild-Str. 1, 85740 Garching, Germany}
\begin{document}


\pagerange{\pageref{firstpage}--\pageref{lastpage}} \pubyear{2004}

\maketitle

\label{firstpage}

\begin{abstract}
Rapid, hyper-Eddington accretion is likely to power the central engines
of  gamma-ray bursts (GRBs). In the extreme conditions of densities
and temperatures the accreting torus is cooled by neutrino
emission rather than by radiation. Another important cooling
mechanism is the advection of energy into the central black hole.
We compute the time evolution of a neutrino-dominated disc that
proceeds during the burst and investigate the changes in its density
and temperature. 
The discrimination between short and long bursts is
made on the basis of the different rates of material inflow to the
outer parts of the disc, thus favoring the binary merger scenario for
the short GRBs and the collapsar scenario for the long ones. 
Within the context of the collapsar model, we also study the evolution of the
photon luminosity of the remnant disc up to times $\sim $ day, and
discuss its implications for the production of emission lines in GRB
spectra.

\end{abstract}

\begin{keywords}
accretion, accretion discs  -- black hole physics 
-- gamma rays: bursts -- neutrinos
\end{keywords}

\section{Introduction}

In the current view, GRBs are thought to be produced in relativistic
ejecta that dissipate energy by internal shocks (Rees \& Meszaros
1994). However, other gamma-ray production mechanisms have also
been proposed; these include magnetic field reconnections (Drenkhahn \&
Spruit 2002; Sikora et al. 2003) in Poynting-flux dominated explosions
(Lyutikov \& Blandford 2003), thermal photospheric emission
(Paczy\'nski 1986; Thompson 1994), Compton drag (Lazzati et al. 2000)
or photon-pair cascades and multiple Compton scatterings (Stern 2003).
The proposed origin of the relativistic ejecta invokes the birth of a
black hole in a catastrophic event. This is thought to take place
after the merger of two neutron stars or a neutron star and a black
hole (Eichler et al. 1989; Paczy\'nski 1991; Narayan, Paczy\'nski \&
Piran 1992) or in a ``collapsar'' (Woosley 1993; Paczy\'nski 1998).
Both these models involve a stage in which the black hole is
surrounded by a debris torus of a large mass, rapidly accreting onto
it. The most efficient cooling mechanism for the dense torus is
provided by neutrino emission, and neutrino annihilation seems to be
an important process for the fireball generation (even though it
occurs on timescales that are typically too short to provide the bulk
of the GRB energy; e.g.  Ruffert \& Janka 1999; 2001 but see also
Setiawan, Ruffert \& Janka 2004).

The durations, $T$, of GRBs range from milliseconds to over a thousand
of seconds, and are distributed in two distinct peaks that constitute two
GRB classes: short ($T\la 2$ sec) and long bursts ($T\ga 2$ sec;
Kouvelietou et al. 1993).  For the long bursts, signatures of an
accompanying supernova explosion have been detected in the afterglow
spectra (Stanek et al. 2003). This strongly favors the ``collapsar''
interpretation for their origin. Furthermore, the GRB positions inferred
from the afterglow observations are consistent with the GRBs being
associated with the star forming regions in their host galaxies.
However, all the observations that we have of the GRBs counterparts,
i.e. afterglows and host galaxies, have been obtained for the long
bursts (see the review by Zhang \& Meszaros 2003) and no such
signatures have been observed in the case of short bursts.
Therefore the merger model is still thought to be viable
for the case of short bursts, the nature of which remains mysterious.

In the merger scenario, the duration of the GRB event is much longer
than the dynamical timescale but rather comparable to the viscous
timescale of the accretion disc. Therefore the phase of accretion sets
the burst duration, differently from the collapsar model where there
is an external reservoir of stellar matter which feeds the accretion
torus. In general, accretion discs in the context of GRBs are expected
to have typical densities of the order of $10^{10-12}$ g cm$^{-3}$ and
temperatures of $10^{11}$ K within $10-20$ Schwarzchild radii ($R_{\rm
  S} = 2GM/c^2$).  Thus, the accretion proceeds with rates of a solar
to several solar masses per second. In this ``hyper-accreting'' regime,
photons become trapped and are not efficient at cooling the disc.
Neutrinos, however, are produced by weak interactions in the very
dense and hot plasma (``neutrino-dominated accretion flow'', NDAF,
Popham Woosley \& Fryer 1999; Narayan, Piran \& Kumar 2001; Kohri \&
Mineshige 2002; Di Matteo, Perna \& Narayan 2002).

In this paper we compute the time-dependent evolution of a massive
accretion disc. We perform our calculations both within the context of
a compact object merger scenario and in the collapsar model, where the
external supply of a matter, e.g. from the fallback from a supernova
explosion (MacFayden, Woosley \& Heger 2001), is essential to extend
the burst duration into the range appropriate for long bursts. We
investigate the evolution of the disc dominated by neutrino cooling,
and study the changes of neutrino and photon luminosities with time.  In
particular, we focus our discussion with the long term evolution (few
tens of hours) of these discs and its implication for emission line
models reported in the early X-ray afterglow of several GRBs. Our work
is thus complementary to the 3D (or 2D) time-dependent numerical
studies of Setiawan et al. (2004) and Lee \& Ramirez-Ruiz (2002) in
being able to extend to the study of the evolution of the bursts (in
both collapsar and merging scenario) to much longer timescale. The
paper is organized as follows. In Section \ref{sec:method} we describe
the basic assumptions of the model of hyper-Eddington accretion disc
and the method used in time-dependent numerical simulations. In
Section \ref{sec:results} we show the evolution of the disc density
and temperature, as well as the resulting neutrino luminosity. In
Section 4 we discuss the implications of the long term evolution of
the accretion disc for models of line production in GRB afterglow
spectra.  Finally, we summarize our results in Section \ref{sec:diss}.

\section[]{Method}
\label{sec:method}

The structure of a steady-state, neutrino-dominated accretion flows
(NDAFs)
has recently  been studied by Popham, Woosley \& Fryer (1999), 
Kohri \& Mineshige (2002) and Di Matteo, Perna \& Narayan (2002).
Here we use a similar prescription to calculate the initial disc
configuration, which afterwards is subject to viscous evolution.

\subsection{Assumptions and model parameters}
\label{sec:assum}

Throughout our calculations we use the vertically integrated equations
of the disc structure, however we note, that in most of the cases the
disc becomes moderately geometrically thick ($H \sim 0.5 r$).  We
assume that the disc extends not further out from the central black
hole than 50 $R_{\rm S}$, as shown e.g. in the recent simulations
performed by Rosswog, Speith \& Wynn (2004). If the disc size were larger, 
the flow
would become highly convection-dominated instead of forming an NDAF
(Narayan, Piran \& Kumar 2001). The initial mass of such a disc is
$M_{\rm disc} \sim 4\pi R^{3}_{\rm out} \Sigma_{\rm out}$, which is
$\approx 0.35 M_{\odot}$ for the accretion rate $\dot M = 1
~M_{\odot}$/s.  The angular velocity of the disc is assumed to be
Keplerian, $\Omega = \sqrt{GM/r^{3}}$, and the sound speed is $c_{\rm
  s}=\sqrt{P/\rho} = \Omega H$. A non-rotating, Schwarzschild black
hole is assumed and the inner radius of the disc is always at 3
$R_{\rm S}$.

For the disc heating we use the standard $\alpha$ viscosity prescription of 
Shakura \& Sunyaev (1973) 
and we adopt a canonical value of $\alpha = 0.1$.
The total pressure $P$ consists of the gas and radiation pressure as
well as the pressure of degenerate electrons:
\begin{equation}
P = P_{\rm rad} + P_{\rm gas} + P_{\rm deg}
\label{eq:ptot}
\end{equation}
\begin{equation}
P_{\rm gas} = {k \over m_{\rm p}}\rho T ({1 \over 4}+{3 \over 4} X_{\rm nuc})
\end{equation}
\begin{equation}
P_{\rm rad} = {11 \over 12}a T^{4}
\end{equation}
\begin{equation}
P_{\rm deg} = {2 \pi h c \over 3} ({3 \over 8 \pi m_{\rm p}})^{4/3}
({\rho \over \mu_{e}})^{4/3}
\end{equation}
where we take the mass fraction of free nucleons $X_{\rm
nuc}=30.97(\rho/10^{10})^{(-3/4)}(T/10^{10})^{(9/8)}\exp(-6.096/T/10^{10})$
if $X_{\rm nuc}<1.0$ and $X_{\rm nuc}=1.0$ elsewhere, and we assume 
the molecular weight per electron $\mu_{\rm e}=2$. The radiation pressure includes
the contribution from the electron-positron pairs via the coefficient 11/12.
 The radiation pressure is always taken into account in the hydrostatic balance and thermodynamic relations, however we neglect it in the viscous heating. 
Therefore we have:
\begin{equation}
Q^{+}_{\rm visc}={3 \over 2}\alpha \Omega H(P_{\rm gas}+ P_{\rm deg}).
\end{equation}

The cooling in the disc is due to advection, radiation and neutrino
emission. The advective cooling in a stationary disc is determined from 
the global ratio of the total
advected flux to the total viscously generated flux (e.g. Paczy\' nski \&
Bisnovatyi-Kogan 1981; Muchotrzeb \& Paczy\' nski 1982; Abramowicz et al. 1988):
\begin{equation}
Q^{-}_{\rm adv}= {F_{\rm adv} \over F_{\rm tot}}=
- {2 r P q_{\rm adv} \over 3 \rho GM}
\label{eq:fadv}
\end{equation}
and
\begin{equation}
q_{\rm adv}=(12-9 \beta) {\partial \ln T \over \partial \ln r} - 
(4-3 \beta){\partial \ln \rho \over \partial \ln r}.
\label{eq:qadv}
\end{equation}
Here $\beta$ is the ratio of the gas plus degeneracy pressure 
to the total pressure
$\beta = (P_{\rm gas}+ P_{\rm deg})/P$. 
In the initial stationary disc we  assume 
that $q_{\rm adv}$ is approximately constant and of order of unity,
but in the subsequent time-dependent evolution the advection term is calculated more
carefully, with appropriate radial derivatives.

In case of photon and electron-positron pairs in the plasma
the radiative cooling is equal to:
\begin{equation}
Q^{-}_{\rm rad}={3 P_{\rm rad} c \over 4\tau}={11 \sigma T^{4} \over 4
\kappa \Sigma}
\label{eq:qrad}
\end{equation}
where we adopt the Rosseland-mean opacity
$\kappa=0.4+0.64\times 10^{23}\rho T^{-3}$ cm$^{2}$/g. 

The most significant neutrino emission processes are: 
$\dot q_{\rm Ne} = 9\times
10^{33}(\rho/10^{10})(T/10^{11})^{6}X_{\rm nuc}$ ergs/cm$^{3}$/s, 
the cooling rate in the non-degenerate URCA process (Qian \& Woosley
1996); $\dot q_{\rm e^{+}e^{-}} = 4.8\times 10^{33} (T/10^{11})^{9}$
ergs/cm$^{3}$/s,  the emission due to the electron-positron pair
annihilation (Itoh et al. 1989); $\dot q_{\rm brems} = 1.5\times
10^{27} (T/10^{11})^{5.5}(\rho/10^{10})^{2}$ ergs/cm$^{3}$/s is the
the bremsstrahlung emission in the non-degeneracy regime of
nucleons (Hannestad \& Raffelt 1998) and $\dot q_{\rm plasmon} = 1.5
\times 10^{32} (T/10^{11})^{9} f(\gamma_{\rm p})$ ergs/cm$^{3}$/s is
the plasmon decay rate (Ruffert, Janka \& Sch\"afer 1996). 
The largest contribution to the neutrino emission 
is due to the electron capture on nucleons $\dot q_{\rm Ne}$. The electron-positron
annihilation rate could be neglected in the case of complete degeneracy of electrons,
i.e. for $\lambda_{\rm e}/kT >> 1$, where $\lambda_{\rm e}$ is the chemical potential 
of electrons. However, in the subsequent calculations the degeneracy parameter is 
never extremely large, and therefore   $\dot q_{\rm e^{+}e^{-}}$ is also taken 
into account.

In the
above processes either electron flavor neutrinos and antineutrinos are
produced (URCA process, plasmon decay), or all the neutrino flavors,
including the heavy lepton neutrinos and antineutrinos (pair
annihilation, bremsstrahlung). In this work we further neglect the
possible anisotropies between neutrinos and antineutrinos as well as
between different flavors, that could arise e.g. from a multiflavour
neutrino leakage scheme (e.g. Rosswog \& Liebend\"orfer 2003).

We treat, in a simplified scheme, neutrino cooling both in the
optically thin and optically thick regimes.

In the (i){\it optically thin regime} the cooling is simply given by:
\begin{equation}
Q^{-}_{\nu}=H(\dot q_{\rm Ne} + \dot q_{\rm e^{+}e^{-}} + \dot q_{\rm
brems} + \dot q_{\rm plasmon}).
\label{eq:qnu}
\end{equation}

In the (ii){\it optically thick regime} we include in our calculations
the prescription for the neutrino opacity according to Di Matteo et
al. (2002).  This will allow us to switch smoothly between the
optically thick and thin regimes. Each neutrino emission process
described above has an inverse process corresponding to absorption.
Absorption onto protons or onto neutrons (the inverse of the URCA
process) gives rise to the large majority of the (a){\it absorptive
  optical depth}, thus given by (see Di Matteo et al. 2002 for
details):
\begin{equation}
\tau_{\rm abs} = {\dot q_{Ne} H \over 4\times (7/8) \sigma T^{4}} =
4.5\times 10^{-7} (T/10^{11})^{2} X_{\rm nuc} (\rho/10^{10}) H.
\label{eq:tau1}
\end{equation}

A very important contribution to the overall opacity for all neutrino
species comes from neutral-scattering off nucleons.
The (b) {\it scattering optical depth} is given by (see Di Matteo et
al. 2002 for derivation of scattering optical depth):
\begin{equation}
\tau_{\rm sc} = 3 \rho \kappa_{s} H = 
8.1\times 10^{-7} (T/10^{11})^{2} (\rho/10^{10}) H
\label{eq:tau2}
\end{equation}
where the factor $3$ assumes equal contribution for all neutrino
species, i.e. thermal equilibrium.

The neutrino cooling in this case is then given by:
\begin{equation}
Q^{-}_{\nu} = { {7 \over 8} \sigma T^{4} \over 
{3 \over 4} \left({ \tau_{\rm abs} + \tau_{\rm sc} \over 2} 
+ {1 \over \sqrt 3} + 
{1 \over 3\tau_{\rm abs}}\right) }.
\label{eq:qnuthick}
\end{equation}
In addition, in this regime, the equation of state should 
contain the neutrino pressure:
\begin{equation}
P_{\nu} = {1 \over 3}\times {7 \over 8} a T^{4} { { \tau_{\rm abs} + \tau_{\rm sc} \over 2} + {1 \over \sqrt 3} \over { \tau_{\rm abs} + \tau_{\rm sc} \over 2} + {1 \over \sqrt 3} + {1 \over 3\tau_{\rm abs}}}
\end{equation}
Finally, the entropy density due to neutrinos, $S_{\nu} = 4/3\times
7/8 a T^{4}$, is included in the advective cooling when we
calculate the initial disc configuration.


The last equation that is essential to close our set of equations
for the stationary disc is
the total energy flux $F_{\rm tot}$. It is dissipated within the disc at a 
radius $r$ and is determined by the global parameters:
\begin{equation}
F_{\rm tot} = {3 G M \dot M \over 8 \pi r^3}. 
\label{eq:ftot}
\end{equation}
(Note, that the above equation will not be used in the subsequent
time-dependent calculations.)
In order to calculate  the initial stationary  configuration, 
we solve the energy balance:
$F_{\rm tot} = Q^{+}_{\rm visc} = Q^{-}_{\rm adv}+Q^{-}_{\rm
  rad}+Q^{-}_{\nu}$. 
In this energy balance we did not include the cooling term due to
the photodisintegration (Di Matteo et al. 2002), while we include
the radiative cooling.
We checked that this term is much less than the 
neutrino cooling rate in the innermost parts of the disc, when the hot disc 
is cooled by neutrinos, and less than advective and radiative cooling, when these two
mechanisms start playing important role.

The initial configuration of the disc is calculated
by means of a simple Newtonian
method. This lets us determine the radial profiles of density and
temperature, as well as the disc thickness and opacity at time $t=0$.

\subsection{Time evolution}
\label{sec:evol}

In the previous subsection we described the initial, stationary disc
configuration. Having computed the initial disc state we allow the
density and temperature to vary with time. We solve the time-dependent
equations of mass and angular momentum conservation for such a disc:
\begin{equation}
{\partial \Sigma \over \partial t}={1 \over r}{\partial \over \partial
r}(3 r^{1/2} {\partial \over \partial r}(r^{1/2} \nu \Sigma))
\end{equation}
and the energy equation:
\begin{eqnarray}
{\partial T \over \partial t} + v_{\rm r}{\partial T \over \partial r}
= {T \over \Sigma}{4-3\beta \over 12-9\beta}({\partial \Sigma \over
\partial t}+  v_{\rm r}{\partial \Sigma \over \partial r}) \\
\nonumber +{T\over P H}{1\over 12-9\beta} (Q^{+}-Q^{-}).
\end{eqnarray}
Here $\Sigma=H \rho$ is the surface density and $v_{\rm r} \approx (3\nu)/(2r)$
is the radial velocity in the disc,
$\nu=(2P\alpha)/(3\rho\Omega)$ is the kinematic viscosity. 
The cooling term $Q^{-}$ consists
of radiative and neutrino cooling, given by Equations \ref{eq:qrad} and
\ref{eq:qnu} (or \ref{eq:qnuthick}).  The advection is included in 
the energy equation
via the radial derivatives.  In the heating term $Q^{+}$ 
we include gas and degeneracy
pressure (in the optically thin regime) and also neutrino pressure 
(in the optically thick regime).
 Note, that this equation includes the radiation entropy, via the coefficient 
$\beta$, which is not equal to unity due to the presence of radiation pressure in the 
total pressure.

We solve the above set of time-dependent equations using the convenient
change of variables, $y=2r^{1/2}$ and $\Xi = y \Sigma$, at the fixed
radial grid, equally spaced in $y$ (see Janiuk et al. 2002 and references therein). 
The number of radial zones is set
to 20. After determining the solutions for the first 100 time steps by the
fourth-order Runge-Kutta method, we use the Adams-Moulton 
predictor-corrector method, allowing the time-step to vary,
when needed. 

We choose the no-torque  inner boundary condition, $\Sigma_{\rm in} =
T_{\rm in} = 0$ (see Abramowicz \& Kato 1989). 
The outer boundary of the disc is parameterized by an
external accretion rate $\dot M_{\rm ext}$ in case of the collapsar scenario.

 In case of the merging compact 
objects scenario there is no matter supply to the disc, and we can have
  $\Sigma_{\rm out} = T_{\rm out} = 0$. However, the outer edge of the disc should 
expand to conserve the angular momentum, and in principle the density would vanish
at the infinite distance.
In order to satisfy this requirement, we supplement our radial grid with
10 empty rings between 50 and 96  $R_{\rm S}$, while the outer boundary condition 
$\Sigma=T=0$ is located at 96 $R_{\rm S}$.
These empty rings gradually fill with material that sinks from the outermost 
disc radius by means of the viscous 
diffusion, and take the angular momentum form the disc.  The extension of the empty 
zone is large enough to ensure that at the end of the calculations  the
surface density in N-1 ring  is less than 5\% of the 
surface density in the zone 20 (the edge of the initial disc).

\section{Results}
\label{sec:results}

We first  analyze the local evolution of the disc on the $T-\Sigma$ plane,
and then we show examples of time evolution of the neutrino luminosity for 
a chosen set of parameters.

\subsection{Stability curves}
\label{sec:stabil}

\begin{figure}
\epsfxsize = 250pt
\epsfbox{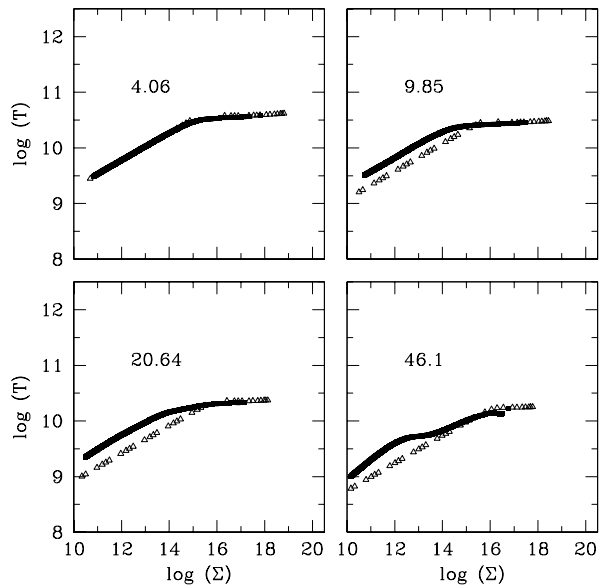}
\caption{Local evolution of the disc on the surface density -
temperature plane, plotted for 4 values of radius: 4.06, 9.85, 20.64
and 46.1 $R_{\rm S}$, in case of the disc optically thin to neutrinos.
 The open triangles mark the stability curve
resulting from stationary disc solutions. The solid points are the 
time-dependent solutions.
\label{fig:ap2}}
\end{figure}

A few examples of stability curves calculated at 4, 10, 20 and 46 
$R_{\rm S}$ for the mass
of the central object equal $3 M_{\odot}$
are shown in Figure \ref{fig:ap2}. The open points represent the
stationary solutions, calculated from the grid of steady-state models
(optically thin regime).
The solid points represent the subsequent solutions of the
time-dependent model.
Deviation between  the stationary and time-dependent solutions 
 arises from the simplified
description of advection, used in the stationary model. In that model 
we had a constant value of the advection parameter given by Eq. \ref{eq:qadv}: 
$q_{adv}=1.0$
and does not depend on radius, while in the time dependent
model the radial derivatives are calculated properly.
However, this simplification of the stationary model does not influence the
time-dependent results, since we start the evolution from the upper, neutrino-cooled
branch, where advection term is not crucial.

We see that the resulting curves display the  pattern already
discussed
in Kohri \& Mineshige (2002). Both upper and lower branches 
of different slopes are viscously and thermally 
stable. On the upper branch the dominant cooling
mechanism is the neutrino cooling, and the radiation pressure is 
much lower than gas and degeneracy pressures (most of this branch is
in fact gas-pressure dominant, and the degeneracy pressure can exceed
it only for very large densities, $\rho > 3\times 10^{12}$ g/cm$^{3}$).
On the lower branch the radiation pressure dominates, however the flow
is cooled by advection rather than radiation, and therefore the disc
should remain stable. Below this branch, for much lower densities and 
temperatures, we expect the unstable (radiatively cooled and
radiation pressure dominant) branch to appear, as shown in detail by 
Pringle, Rees \& Pacholczyk (1974) and Lightman \& Eardley (1974). 

We start the disc evolution from the top of the upper branch, and
first we let it cool down very quickly due to the zero mass inflow at
the outer edge. In the initial configuration, neutrino cooling is
dominant in the entire disc. Although the outer boundary condition
imposes a fast density decrease at the outer edge of the disc,
 the inner parts do not respond immediately to that.
For the first $\sim 1$ second the disc annuli evolve
slowly along the upper branch, and the density of the disc decreases
as the matter falls into the black hole. In this branch the
temperature changes are not very significant, so it is the substantial
density evolution that drives the steep fall of the disc luminosity.

The evolution starts to proceed faster when all the disc annuli
find themselves on the lower branch, where advective cooling becomes
dominant. 
 As the whole
disc is dominated by advective cooling,
we reach the so-called slim disc branch. Here the radiation pressure 
is larger than gas and degeneracy pressure.
 As a result of this, we may expect the
outer disc edge to develop an instability leading to the disc break
down,
 if the viscous heating
was assumed proportional to the total (gas and radiation) pressure.
 On the other hand, if the viscous heating is assumed proportional only 
to the gas pressure,
there is no problem with instabilities during  the slim disc evoluton. 
Therefore we choose this prescription,  also having in mind the recent results
of 3-D simulations performedd by Turner (2004), which
show that the radiation pressure instability is strongly suppressed when the
angular momentum transport in the accretion disc is described by MHD calculations,
without the $\alpha$ parameter. 
This heating prescription makes no  difference at the NDAF branch, 
where more important 
contribution is given by gas, degeneracy and neutrino pressure, which we always
take into account.
Since  neutrino cooling is no longer important
($L_{\nu} < 10^{50}$ ergs/s), further calculations would not be relevant
for the the study of GRB central engine and we stop our calculations here.

The disc evolution proceeds much slower when the accretion rate at the
outer edge is non-zero. The phase of the most efficient neutrino
cooling, i.e. when the innermost disc parts are cooled mainly by the
neutrino emission and not by advection, is extended to $\sim 10$
seconds. Also the second phase, of the advection-dominated disc, is
much longer. The disc evolves along the advective branch more slowly
and there is no rapid emptying of the outer parts.

\subsection{Lightcurves}
\label{sec:lcurves}

The outer boundary condition for the disc evolution depends on the
accretion rate, $\dot M_{\rm ext}$. If we assume that the torus was
formed during the merger of two compact objects, NS-NS or BH-NS, it
will consist of the debris matter from the disrupted neutron star and
there will be no source of additional material infalling onto
the disc. On the other hand, if we take into account the ``collapsar''
model, the associated supernova explosion should provide a declining
fallback of matter, that accumulates mostly in the equatorial plane.
The variations of the mass supply to the accretion flow will strongly
influence the disc evolution and therefore result in significant
changes in the lightcurves.

The lightcurves represent the neutrino luminosity of the disc.
The luminosity is given by:
\begin{equation}
L_{\nu}=\int_{R_{\rm min}}^{R_{\rm max}} Q^{-}_{\nu} 2\pi r dr
\label{eq:lumthin}
\end{equation} 
where $Q^{-}$ is given by either Equation \ref{eq:qnu} or \ref{eq:qnuthick}
 as appropriate
for the optically-thin and thick regimes.

\subsubsection{Optically thin regime}
\label{sec:opthin}

\begin{figure}
\epsfxsize = 250pt
\epsfbox{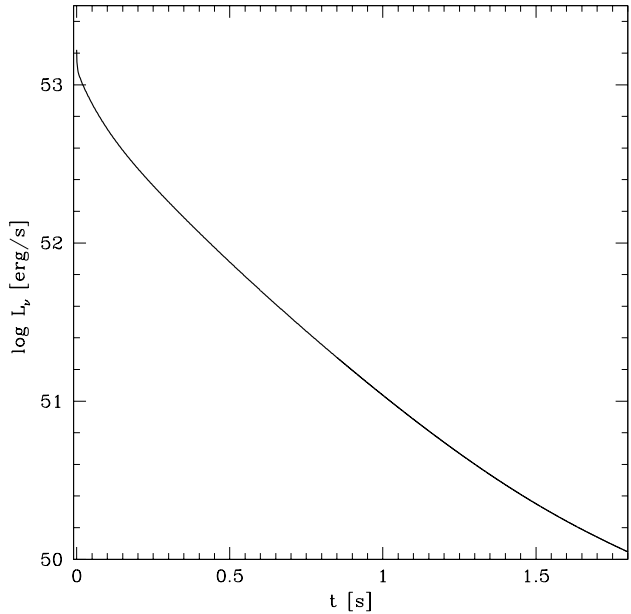}
\caption{The time evolution of the neutrino luminosity in case of
no external mass supply to the disc (optically thin case). 
The initial accretion rate is
$\dot M=1~ M_{\odot}$/s.
\label{fig:alfa}}
\end{figure}

In Figure \ref{fig:alfa} we show an example of a neutrino lightcurve
calculated under the assumption of an optically-thin neutrino disc for the case
of zero mass supply.
Other parameters were always the same: mass of the central object $M=3M_{\odot}$, 
$\alpha = 0.1$, $R_{\rm out} = 50 R_{\rm S}$, and the starting accretion rate was $\dot M_{0}=1 ~M_{\odot}$/s.

\begin{figure}
\epsfxsize = 250pt
\epsfbox{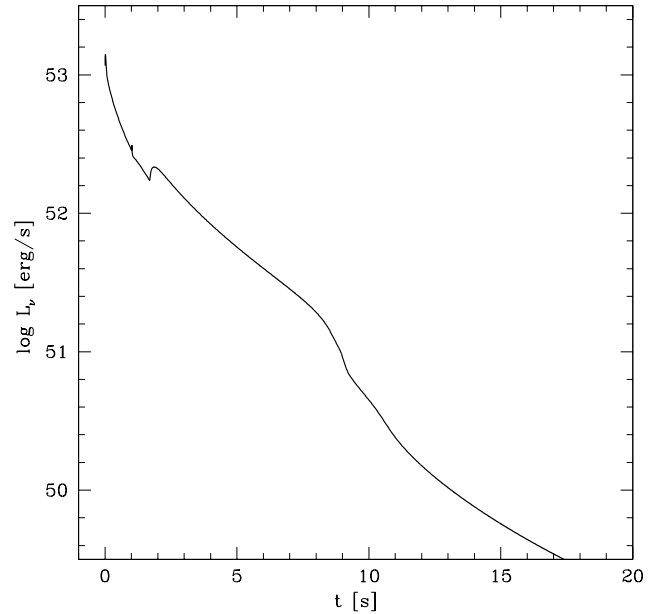}
\caption{The time evolution of the neutrino luminosity in case of
additional external mass supply to the disc, declining with time as in
the canonical supernova
fallback prescription, $\dot M_{\rm ext} \sim t^{-5/3}$. 
The initial accretion rate is
$\dot M=1~ M_{\odot}$/s. The disc is optically thin to neutrinos.
\label{fig:beta}}
\end{figure}

In Figure \ref{fig:beta} we show the neutrino lightcurve calculated
for the case of additional matter supply at the outer edge of the
disc, which declines with time as:
\begin{equation}
\dot M_{\rm ext} = \dot M_{0} t^{-5/3}
\label{eq:fallb}
\end{equation}
This is the time dependence of the fallback of material derived for the supernova SN 1987A 
(Chevalier 1989) 
and we adopt it as a canonical rate of the fallback in case of GRBs
associated with supernovae. 
The starting accretion rate was again $\dot M_{0}=1 ~M_{\odot}$/s.

Clearly, the supply of material at the outer edge of the disc extends
the duration of the hot phase, when the neutrino cooling dominates.
GRBs associated with a supernova explosion can therefore have much
longer durations, of the order of tens of seconds, depending on the
initial mass of the disc and accretion rate. The neutrino luminosity
declines exponentially, with some slight luminosity fluctuations in the
initial phase, when the neutrino cooling remains dominant at least in
the innermost disc parts. The lightcurve smoothens as soon as the
whole disc switches to the advection-dominated mode, but its slope
does not change much.

\subsubsection{Optically thick regime}
\label{sec:opthick}

When the disc is optically thin to its neutrino emission, the cooling
rate due to neutrinos, $Q^{-}_{\nu}$, is given by Equation \ref{eq:qnu}.
However, the disc can become optically thick to neutrinos, especially
in its innermost regions and for high accretion rates.  The optical
depth is larger than unity in the initial disc state. Although it drops very
fast in the outer disc radii, still remains quite high for most of
the disc evolution in the innermost parts. In this case the
cooling rate due to neutrinos $Q^{-}_{\nu}$ is given by Equation
\ref{eq:qnuthick}.

\begin{figure}
\epsfxsize = 250pt
\epsfbox{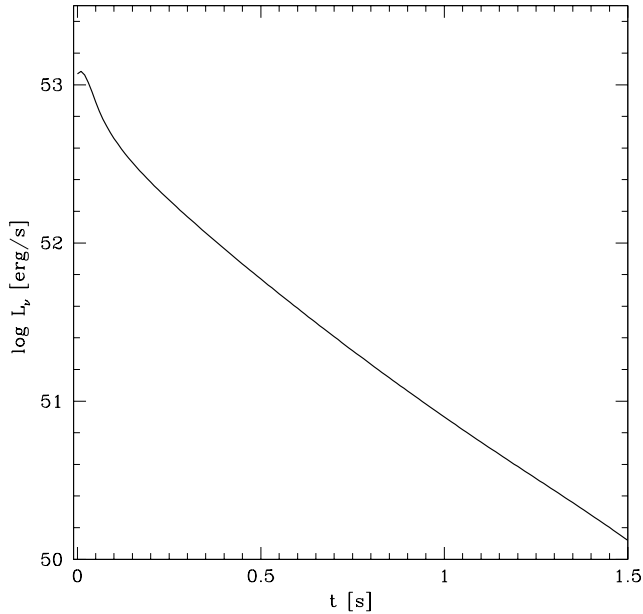}
\caption{The time evolution of the neutrino luminosity in case of
no external mass supply to the disc and neutrino optical depth included in the 
model. 
The initial accretion rate is
$\dot M=1~ M_{\odot}$/s.
\label{fig:alfa-tau}}
\end{figure}

In Figure \ref{fig:alfa-tau} we show the neutrino luminosity resulting
from the model in which we include the treatment of neutrino optical
depths and add the neutrino pressure term in the equations. As
expected, the most significant change in comparison with the optically
thin case (Figure 2) occurs at the very beginning of the evolution,
when the innermost disc is optically thick to its neutrino emission.
After about 0.2 seconds most of the disc becomes optically thin to
neutrinos and the solution is almost the same as in Section
\ref{sec:opthin}.

\begin{figure}
\epsfxsize = 250pt
\epsfbox{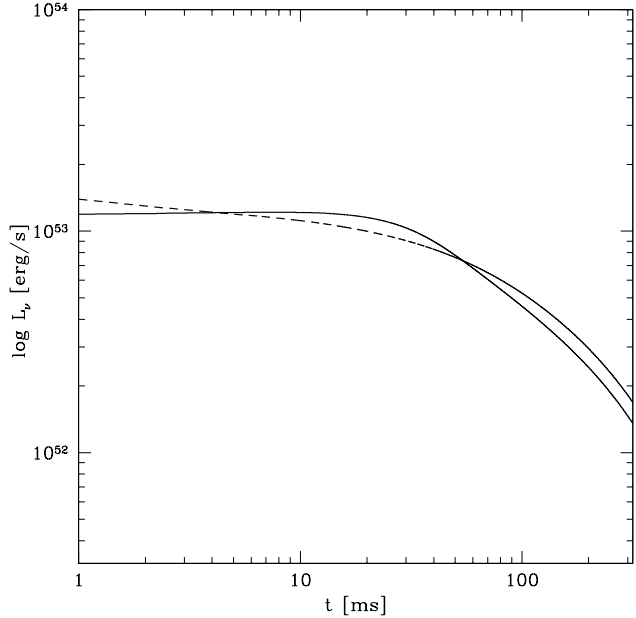}
\caption{The time evolution of the neutrino luminosity in case of
no external mass supply to the disc (compact objects merging scenario). 
The solid line shows the model with neutrino 
optical depth, while the dashed line is the optically thin case.
The initial accretion rate is
$\dot M=1~ M_{\odot}$/s and $\alpha=0.1$.
\label{fig:mili}}
\end{figure}

Recently, Lee et al. (2004) computed the dynamical simulation of a
short GRB, resulting from the merger of two neutron stars. They showed
that inclusion of the neutrino optical depth influences the duration
and variability of the short burst. Our simulation confirms this
conclusion: in Figure \ref{fig:mili} we show the comparison of the
neutrino luminosity evolution in the optically thin and thick regimes,
in the zoom-in emphasizing the shortest timescale by means of the
logarithmic units. As the figure shows, in the optically thick case
the neutrino luminosity around 20 milliseconds is enhanced,
consistently with Lee et al. simulations. Then, the luminosity
declines slightly steeper than for the thin disc and the duration of
the burst is shorter.

\begin{figure}
\epsfxsize = 250pt
\epsfbox{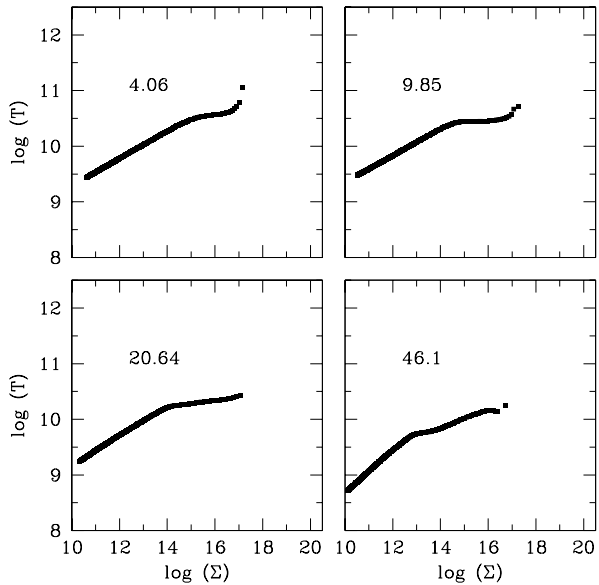}
\caption{Local evolution of the disc on the surface density -
temperature plane, plotted for 4 values of radius: 4.06, 9.85, 20.64
and 46.1 $R_{\rm S}$, in case of the neutrino-thick inner disc.
The points are the 
time-dependent solutions, for the time range 0 - 0.4 seconds.
\label{fig:sttau}}
\end{figure}

In Figure \ref{fig:sttau} we plot the disc evolution on the surface
density vs. temperature plane in case of the disc optically thick to
neutrinos.  The main difference in comparison to the optically thin
case (Figure {\ref{fig:ap2}) is at the inner radii, where the local
  solutions achieve higher temperatures, $T> 10^{11} K$.  Further out,
  and also later in time, when the disc becomes advective and neutrino
  optical depth is less than 1.0, the solutions match the optically
  thin case.

In case of a long burst, the difference between the optically thin and
optically thick regimes is not significant, since
after $\sim$ 1 second of the disc evolution the both optical depths
  $\tau_{\rm abs}$ and $\tau_{\rm sc}$ become less than 1.0 and we can
  substitute the optically thick equations with the optically thin
  ones.  Thus, in practice, the resulting lightcurve does not differ
  from that shown in Figure \ref{fig:beta}.

\section{Long-term evolution of the accretion disc and its implications
for emission line models}
 
Our time-dependent code allows us to follow the evolution of the GRB
accretion disc up to times much longer than the duration of the GRB
itself.  This is valuable information in light of the peculiarities
observed in the early stages of some afterglows (e.g. Fox et
al. 2003), which hint at a more extended period of energy injection
than the GRB phase itself.  What is however particularly relevant for,
is the interpretation of the narrow emission features that have been
reported in the early X-ray afterglow of several bursts (Yoshida et
al. 1999; Antonelli et al. 2000; Piro et al. 1999, 2000; Reeves et al.
2002). Although the statistical significance of the detections is not
compelling, the interpretation of these lines bears important
implications for our understanding of GRB progenitors.

The emission lines are detected at the rest frame frequency of the
host galaxy, and therefore they cannot be produced within the fireball,
which, during the X-ray afterglow phase, is still moving towards the
observer at a very high speed, with a Lorentz factor $\Gamma\ga 10$.
Any emission model for the lines must be confronted (as a minimum
requirement) with the energetics of the lines, which have a luminosity
(for isotropic emission) $\sim 10^{44}-10^{45}$ erg/s and last for
several hours.  This luminosity values must be augmented by a factor
$1/\eta$, where $\eta$ is the efficiency for the reprocessing of the
impinging flux into line photons.  The efficiency of conversion of the
X-ray ionizing continuum into K$_\alpha$ line photons was investigated
by Lazzati et al.  (2002). They assumed that the line is produced by
reflection off an optically thick slab of material, illuminated by a
powerlaw continuum.  Under these conditions, which provide the most
efficient way of reprocessing continuum photons into K$_\alpha$ lines,
they found that the efficiency $\eta$ can be at most 2\%. This implies
that the impinging continuum flux must carry an energy $\ga
10^{46}-10^{47}$ erg/s. 

This energy requirement could in principle be
alleviated if the line emission were collimated. In the model of Rees
\& Meszaros (2000), where lines are produced by multiple scattering
off the walls of an evacuated funnel along the rotation axis of the
collapsing star, there can indeed be some collimation of
the emission, enhancing its observed flux. The importance of this
effect was however shown to be not too significant by Ghisellini et al. (2002). 
In particular, they found that, if the main scattering opacity
within the funnel is provided by free electrons, then photons scattered
deeper in the funnel will be lost, and the resulting line amplification
can be at most a factor of 2. On the other hand, if the line under
consideration is a resonant line, these photons may be rescattered
by ions in the funnel wall and redirected towards the open space. 
In this case, line amplification can
get up to a factor of 10 for an opening angle of the funnel of $\sim 5$ deg
(and it gets smaller for larger angles).
Therefore, even under the most favorable circumstances, the required
luminosity of the continuum has to be  $\ga 10^{45}-10^{46}$ erg/s. 
The issue therefore remains what is that provides the energy to power
the lines.  

\begin{figure}
\epsfxsize = 250pt
\epsfbox{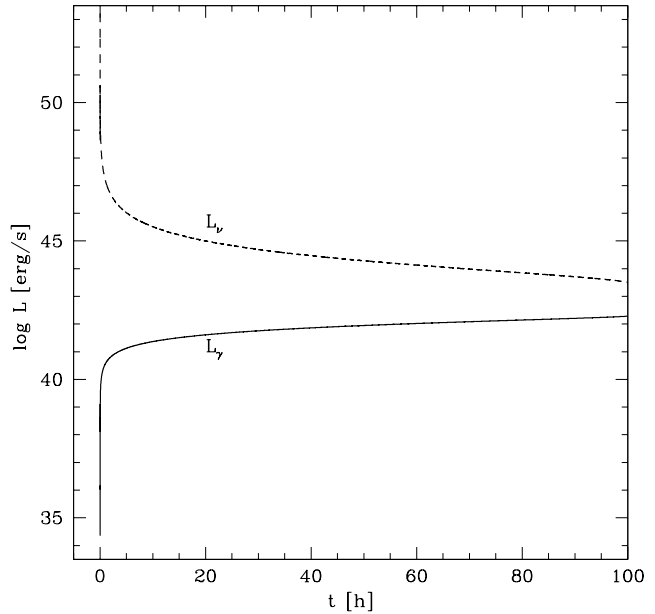}
\caption{Long-term evolution of photon (solid line) 
and neutrino (dashed line) luminosities
of the remnant torus, fed by the material fallback at its outer edge
as in Eq. \ref{eq:fallb}.
\label{fig:long}}
\end{figure}

The recent detection of a supernova associated with GRB 030329 (Stanek
et al. 2003; Hjorth et al. 2003) gave strong support to the ideas that
GRBs are the result of explosion of massive stars. In the collapsar
model (MacFadyen \& Woosley 1999), the remnant compact object is a
black hole, and the ``engine'' is an accretion disc like the one we
have studied here. The energy output at later times (i.e. after the
GRB phase) is in this case given by the sluggish drain of orbiting
matter into the newly formed black hole, after the supply of matter
from the envelope of the collapsing star is exhausted.  Our
calculations allow us to evaluate the energy output from this remnant
disc at the times (of the order of tens of hours) at which the
emission features have been observed.  Figure \ref{fig:long}
 shows the evolution
of the photon and neutrino luminosities from the very early times of
the GRB phase to several days after the GRB. While the neutrino
luminosity drops following the decrease in the accretion rate 
(Figure \ref{fig:mdot}), the photon luminosity increases with time. 
This is due to
the decreasing photon opacity of the disc,
which overcomes
the decrease in energy flux that would be expected as a result of the
declining mass accretion rate. 
The Rosseland mean opacity, calculated for the electron scattering and 
free-free transitions (see Equation \ref{eq:qrad}) is very large
in the initial disc, which is therefore almost totally opaque to photons
($\kappa > 3\times10^{2}$ for r $< 10 R_{\rm S}$). However, the opacity 
drops quickly and after $\sim 10$ minutes it approaches the value of 0.4.
Therefore as long as the main source of cooling is the neutrino emission,
the photon luminosity rises with time.
The trend will reverse when $L_{\nu} < L_{\gamma}$ and the photon emission 
becomes the more important cooling mechanism (note however that advection is
still very large). The decreasing mass accretion rate will now
result in lower and lower photon emission from the disc.

As the figure shows, at the times
corresponding to the line observations, the photon luminosity is
$\sim 10^{41}-10^{42}$ erg/sec, several orders of magnitude below what 
is required to produce the emission features at the observed luminosity level.
The simulations in Figures \ref{fig:long} and \ref{fig:mdot} 
are performed for an initial accretion rate of 1 $M_{\odot}/s$, but they 
give similar results for any initial value of $\dot{M}$ that produce a GRB at early
times. Since for higher initial accretion rates the neutrino cooling of the disc
is even more effective, the photon luminosity at late times is lower:
for $2 M_{\odot}$/s it is $3.2\times 10^{41}$ erg/s at $t\sim 1$ day, while
for $1 M_{\odot}$/s it was $4.7\times 10^{41}$ erg/s.
The viscosity in the disc does not have a big impact on the 
long-term results either: the dissipation rate and 
the disc luminosity increase only slightly with $\alpha$
and the change is of the order of $\Delta \log L_{\gamma} = 0.3$
and $\Delta \log L_{\nu} = 0.5$ when we change $\alpha$ form 0.1 to 0.3. 
Therefore, our simulations imply that the
energy output from the accretion disc alone is not sufficient to power the 
GRB emission lines (if these are indeed real).

\begin{figure}
\epsfxsize = 250pt
\epsfbox{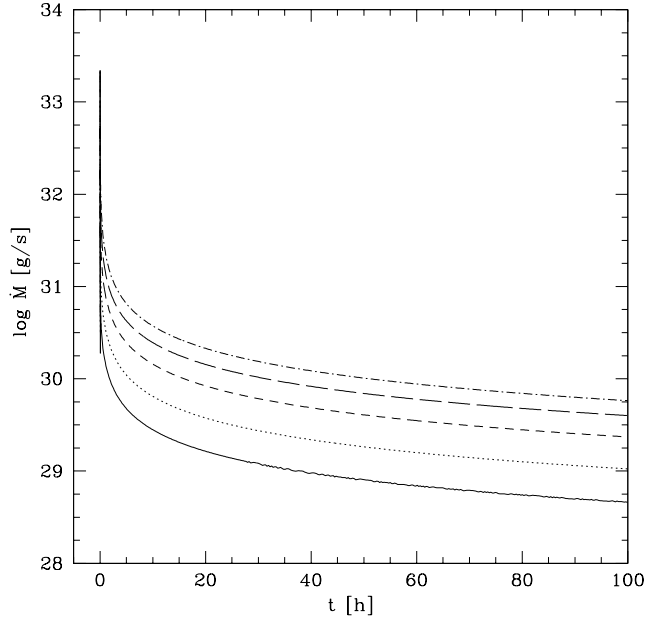}
\caption{Long-term evolution of the accretion rate in the torus, 
plotted for several radii: 5.27 $R_{\rm S}$ (solid line),
9.85 $R_{\rm S}$ (dotted line), 20.64 $R_{\rm S}$ (short-dashed line),
35.37 $R_{\rm S}$ (long-dashed line) and 46.1 $R_{\rm S}$ (dot-dashed line).
\label{fig:mdot}}
\end{figure}

An alternative possibility (Rees \& Meszaros 2000) for producing a
long-lasting energy output is by the collapse of a massive star into a
fast millisecond pulsar (Usov 1994; Thompson 1994; Wheeler et al. 2000).
In this case the late time energy is provided by the
electromagnetic losses of the pulsar which, at a day after the burst,
can easily be as high as $10^{47}$ erg/s.  
Our results favor this scenario. Identification of the
remnant compact object after the GRB explosion would provide an
independent test of this scenario, and this should be possible in
local galaxies after identification of GRB remnants (Loeb \& Perna
1998; Efremov et al. 1998; Perna et al. 2000).

\section{Discussion}
\label{sec:diss}

The bimodal distribution of durations of GRBs is most probably 
due to the different origin of the short and long bursts, and 
the proposed models of a GRB progenitor include binary mergers
or collapse of a massive star. In both cases the formation of a dense
 accretion disc around a black hole is expected, and one of the possible ways
of extracting the energy from the central engine is the neutrino-antineutrino
annihilation. This process can give rise to the relativistic fireball, 
if the neutrinos convert into electron-positron pairs in the region of
relatively low baryon density. 

The merger scenario seems to be appropriate only in the case of short
GRBs, of duration $\la 2$ seconds. Due to the need for a low baryon
density above the poles, the favored case is black hole - neutron star merger.
The neutron star can  either be catastrophically disrupted in the first
approach to the black hole or transfer mass to the BH 2 or 3 times, depending on 
the initial mass ratio. In either case in the end it
forms a torus, whose mass is between 0.25 and 0.7 $M_{\odot}$, 
and accretion rate of several $M_{\odot}$/s. 
Therefore the accretion time is of order of 0.1 - 1.5 sec
(Janka et al. 1999).

In the collapsar scenario the black hole formation may be accompanied by 
a fallback of material, which over a period of minutes to hours feeds the
accretion disc. If the initial accretion rate is sufficiently high to establish
a dense disc cooled by neutrinos ($\dot M \ge 0.1 ~M_{\odot}$/s), the  
powerful, long GRB may be produced. 

The neutrino annihilation is most efficient near the rotation axis (Jaroszy\'nski 1993). 
This process is able to account for the energy released in GRBs if a moderate beaming,
$\Omega/2\pi \sim 10^{-2}$, is involved (Ruffert \& Janka 1999). 
In this case the isotropised energy of a burst is of the order of 10$^{51}$-10$^{52}$ 
erg/s. 
The jet collimation mechanism is possible e.g. due to the neutrino-driven wind,
which can be expected from the outer disc regions
if the neutrino luminosity is $L_{\nu} > 10^{52}$ ergs/s and the 
 estimated wind power is 
$L_{\rm w} = 2 \times 10^{49} (L_{\nu}/10^{52})^{(16/5)}$ (Rosswog \& Ramirez-Ruiz 2003). 

In addition to the fraction of accretion energy that goes into the jet via neutrinos,
the  energy extraction from the central engine is possible via magnetic fields 
(Blandford \& Znajek 1977; Meszaros \& Rees 1997; Klu\'zniak \& Ruderman 1998).
 In this case, isotropic luminosities as high as 10$^{52}$-10$^{53}$ erg/s 
are possible, even without beaming.
The burst energy can also be increased due to the effects associated with the 
Kerr metric of a rotating black hole.

 Here we start our calculations when the accretion disc hass already established 
its quasi-steady state and evolves in the viscous timescale.
We neglect the first, very violent stage of the disc formation, during which
the material within a free fall time settles on the equatorial plane.
The centrifugal force is balanced by gravity and the torus forms with a 
surrounding accretion shock.  Inside this shock the temperatures and densities are
high enough to provide anhanced neutrino emission during a few seconds before the disc 
formation (MacFadyen \& Woosley 1999).

We computed the time dependent evolution of the accretion disc believed
to be the GRB central engine. Our simulations took into account a physical equation
of state, including gas and radiation pressure as well as the pressure from degenerate 
electrons. We calculated the energy balance between the viscous heating and neutrino losses
due to thermal processes and neutronization, in both optically thin and thick regimes.
 We also took into account the
advective and radiative cooling.
Time-dependent 2D hydrodynamical simulations of an accretion disc in
GRBs have recently been calculated by Lee \& Ramirez-Ruiz
(2002), Lee et al. (2004) and Setiawan et al. (2004). 
Here we use the vertically integrated equations and 
the simulations are one-dimensional,
which on one hand results in a great simplification of the problem, 
but on the other hand allows us to get insight into the
various physical processes that act during the disc evolution. The most valuable
aspect of such an approach is the possibility to study the long-term evolution of
the remnant disc, especially in the frame of a long burst (collapsar) scenario.

The short bursts are powered by the neutrino cooled accretion disc, which
exists for about $\sim  2$ seconds. 
At this phase the neutrino luminosity remains 
very high during the first $\sim 1 - 1.5$ seconds, 
while dropping as soon as the whole disc becomes advection dominated.
In case of the long bursts, the fallback of material that occurs during the
accompanying supernova explosion plays an essential role. 
The neutrino luminosity remains very high during the first
2-3 seconds of the burst, and then drops exponentially, being sufficiently high to
feed the GRB during several tens of seconds. 

The hyper-accreting, neutrino cooled disc is in either case very opaque to photons
and the photon luminosity is always negligible whenever the neutrino emission 
is significant. However, as the disc evolves and the neutrino cooling becomes 
less efficient, the disc opacity due to electron scattering and free-free transitions
decreases, thus allowing more and more photons to escape. 
Therefore the photon luminosity rises in time up to the level at which it exceeds 
the decreasing neutrino luminosity. Later on, since the disc is cooled mainly by 
photons (and advection), and during its evolution the accretion rate and energy 
generation decrease, the photon luminosity drops with time.

This stage of significant photon emission from the disc is extremely short and 
therefore not very important in case of the binary merger model.
However, in the collapsar scenario it is much more extended in time and can last
over several days to weeks, depending on the initial accretion rate and viscosity
(the greater initial $\dot M$ and $\alpha$, the longer the photon emission phase).
Our calculations allow us to determine the residual level 
of photon luminosity from the disc at the times ($\sim$ tens of hours) corresponding 
to the claimed observations of X-ray emission features in a few GRBs.

The emission from the disc could in principle power the emission
lines, if the luminosity were sufficiently high. However, in our model
this would require quite low initial accretion rates and an extremely
high value of the viscosity parameter in the disc (for $\alpha > 0.9$
and $\dot M = 0.1 M_{\odot}$/s we could expect the X-ray luminosity of
the order of $10^{44}$ erg/s), whereas for a standard value of
viscosity $\alpha=0.1$, which is commonly adopted, the disc alone is not capable to
power the lines. Therefore our simulations would rather
favor the possibility of a long-lasting energy release by the
electromagnetic losses of a newly born millisecond pulsar.

\section*{Acknowledgments}
We thank  Susumu Inoue, 
Andrea Merloni, Thomas Janka, Marek Sikora and Tomek Bulik for helpful discussions.
This work was supported in part by the Gamma Ray Burst Training Network and by 
grants 2P03D00322 and PBZ 057/P03/2001 of the Polish State Committee for Scientific Research.

\end{document}